\documentclass{llncs}

\usepackage{amsmath,amscd,amssymb,amsfonts,latexsym}
\newcommand{\struct}[1]{\langle #1 \rangle}

\renewcommand{\d}{\delta}
\newcommand{\e}{\varepsilon}

\newcommand{\g}{\gamma}

\newcommand{\s}{\sigma}

\renewcommand{\S}{\Sigma}

\newcommand{\Aa}{{\cal A}}
\newcommand{\Bb}{{\cal B}}

\newcommand{\Tt}{{\cal T}}

\newcommand{\notinconf}[1]{}   
\newcommand{\notinfull}[1]{}   

\newcommand{\clo}{{\cal C}}
\newcommand{\constr}{{\Phi(\clo)}}
\newcommand{\x}{{x}}

\newcommand{\deno}[1]{\lbrack #1 \rbrack}
\newcommand{\rplus}{{\mathbb R}_+}
\newcommand{\powers}[1]{{\cal P
}(#1)}
\newcommand{\ttt}{\mathit{tt}}
\newcommand{\autom}{{\cal A}}
\newcommand{\mach}{{\cal M}}
\newcommand{\bautom}{{\cal B}}
\newcommand{\boo}[1]{{\cal B}^+(#1)}

\newcommand{\tr}[4]{{#1, #2, #3 \ \mapsto \ #4}}
\newcommand{\treq}[4]{#1, #2, #3 & \ \mapsto \  & #4}
\newcommand{\pair}[2]{(#1,#2)}
\newcommand{\pairnb}[2]{{#1,#2}}
\newcommand{\gamepar}[2]{{G_{#1,#2}}}
\newcommand{\game}{\gamepar{\autom}{w}}
\newcommand{\val}{{\mathbf v}}
\newcommand{\zero}{\val_0}
\newcommand{\conf}[2]{\pair{#1}{#2}}
\newcommand{\adam}{Adam}
\newcommand{\ewa}{Eve}
\newcommand{\incr}[2]{{{#1} {+} {#2}}}
\newcommand{\reset}[2]{{{#1}[#2 := 0]}}
\newcommand{\langpar}[1]{{L(#1)}}
\newcommand{\lang}{\langpar{\autom}}
\newcommand{\compl}[1]{{\neg{#1}}}
\newcommand{\syst}{{\cal S}}
\newcommand{\wri}[1]{{!#1}}
\newcommand{\rea}[1]{{?#1}}
\newcommand{\trperf}[3]{{#1}\stackrel{#2}{\longrightarrow}{#3}}
\newcommand{\trlossy}[3]{{#1}\stackrel{#2}{\Longrightarrow}{#3}}
\newcommand{\subse}{\sqsubseteq}

\newcommand{\Sig}{{\overline{\Sigma}}}
\newcommand{\SigH}{{{\Sigma}_{\hea}}}
\newcommand{\ti}{{u}}
\newcommand{\incone}[1]{{(#1 + 1)}}
\newcommand{\automdelta}{{\autom_{\text{strict}}}}
\newcommand{\automstruct}{{\autom_{\text{struct}}}}
\newcommand{\automunit}{{\autom_{\text{unit}}}}
\newcommand{\automcheck}{{\autom_{\text{check}}}}
\newcommand{\qrule}[1]{{s_{#1}}}
\newcommand{\qal}[1]{{\bar{s}_{#1}}}
\newcommand{\qste}{{s_{\text{step}}}}
\newcommand{\qchannel}{{s_{\text{channel}}}}
\newcommand{\qpone}[1]{{s^{+1}_{#1}}}
\newcommand{\qwri}[1]{{s_{\wri{#1}}}}
\newcommand{\qrea}[1]{{s_{\rea{#1}}}}
\newcommand{\qtryrea}[1]{{s_{\text{try} \rea{#1}}}}
\newcommand{\qchrea}[1]{{s_{\text{check} \rea{#1}}}}
\newcommand{\TT}{{\widehat{\cal T}}}
\newcommand{\HH}{{\cal H}}
\newcommand{\tran}[3]{{#1}\stackrel{#2}{\longrightarrow}{#3}}
\newcommand{\next}[1]{{\text{\tt Next}_{\conf{q}{\val_{#1}}}}}
\newcommand{\cmax}{{c_{\text{max}}}}
\newcommand{\reg}{{\text{\tt reg}}}
\newcommand{\creg}[1]{{\text{\tt r}}_{#1}}
\newcommand{\rreg}{{\text{\tt r}}}
\newcommand{\alp}{{\Lambda}}
\newcommand{\fract}[1]{{\text{\tt fract}(#1)}}
\newcommand{\equ}{{\sim}}
\newcommand{\wqo}{\preceq}

\newcommand{\automeps}{{\autom_\epsilon}}
\newcommand{\Sigeps}{{\Sigma_\epsilon}}
\newcommand{\remeps}[1]{{|#1|_\epsilon}}
\newcommand{\automdru}{{\cal A}'}
\newcommand{\automextra}{\widehat{{\cal A}}}
\newcommand{\qone}{{s^{+1}}}
\newcommand{\qacc}{q_{\text{acc}}}
\newcommand{\bla}{\mathtt{B}}
\newcommand{\hea}{\mathtt{H}}
\newcommand{\tomove}[2]{\longrightarrow{#1}{#2}}
\newcommand{\move}[4]{{{#1}{#2}\tomove{#3}{#4}}}
\newcommand{\timedmove}[4]{{{#1}{#2}\leadsto{#3}{#4}}}
\newcommand{\untime}[1]{{\mathtt{untime}}(#1)}
\newcommand{\qleft}[1]{{s^{#1}_{\leftarrow}}}
\newcommand{\qright}[1]{{s^{#1}_{\rightarrow}}}
\newcommand{\qstay}[1]{{s^{#1}_{\cdot}}}
\newcommand{\qstaycont}[1]{{s^{#1}_{\text{cont}}}}

\newcommand{\removed}[1]{}

\begin{document}

\pagestyle{plain}
\pagenumbering{arabic}

%
\title{Alternating Timed Automata
\thanks{Work reported here has been partially supported by the
European Community
Research Training Network {\sc Games}.}}
%
%
\author{S{\l}awomir Lasota\inst{1}\thanks{Partially supported by the
Polish
{\sc Kbn} grant No.\ 4 T11C 042 25.
This work was performed in part during the author's stay
at LaBRI, Universit\'{e} Bordeaux-1.} \and Igor Walukiewicz\inst{2}}
%
%
%
\institute{Institute of Informatics, Warsaw University\\
Banacha 2, 02-097 Warszawa\\
\ \\
\and
LaBRI, Universit\'e Bordeaux-1\\
351, Cours de la Lib\'eration,
F-33 405, Talence cedex, France}

\maketitle              

\begin{abstract}
    A notion of alternating timed automata is proposed. It is shown
    that such automata with only one clock have decidable emptiness
    problem over finite words. This gives a new class of timed
languages which is closed
    under boolean operations and which has an effective presentation. We
    prove that the complexity of the emptiness problem for alternating
    timed automata with one clock is non-primitive recursive. The proof
    gives also the same lower bound for the universality problem for
    nondeterministic timed automata with one clock.
    We investigate extension of the model with epsilon-transitions and
    prove that emptiness is undecidable.
    Over infinite words, we show undecidability of the universality  
problem.
%
\end{abstract}

\section{Introduction}

Timed automata is a widely studied model of real-time systems. It is
obtained from finite nondeterministic automata by adding clocks which
can be reset and whose values can be compared with constants. In this
paper we consider alternating version of timed automata obtained by
introducing universal transitions in the same way as it is done for
standard nondeterministic automata. From the results of Alur and
Dill~\cite{AluDil94} it follows that such a model cannot have
decidable emptiness problem as the universality problem for timed
automata is not decidable.  In the recent paper~\cite{OW04} Ouaknine
and Worrell has shown that the universality problem is decidable for
nondeterministic automata with one clock, over finite timed words.
Inspired by their
construction, we show that the emptiness problem for alternating timed
automata with one clock is decidable as well. We also prove not
primitive recursive lower bound for the problem. The proof implies the
same bound for the universality problem for nondeterministic timed
automata with one clock, thereby answering the question posed by
Ouaknine
and Worrell~\cite{OW04}. To complete the picture we also show that an
extension of our model with $\epsilon$-transitions has undecidable
emptiness problem.
Furthermore, we prove undecidability of the universality problem for
one-clock nondeterministic automata over infinite timed words.

\looseness=-1
The crucial property of timed automata models is the decidability of
the emptiness problem. The drawback of the model is that the class of
languages recognized by timed automata is not closed under complement
and the universality question is undecidable
($\Pi^1_1$-hard)~\cite{AluDil94}. One solution to this problem is to
restrict to deterministic timed automata. Another, is to restrict the
reset operation; this gives the event-clock automata
model~\cite{AluFixHen97}. A different ad-hoc solution could be to take
the boolean closure of the languages recognized by timed automata.
This solution does not seem promising due to the complexity of the
universality problem. This consideration leads to the idea of using
automata with one clock for which the universality problem is
decidable. The obtained class of alternating timed automata is by
definition closed under boolean operations.  Moreover, using the method
of Ouaknine and Worrell, we can show that the class has decidable
emptiness problem.  As it can be expected, there are languages
recognizable by timed automata that are not recognizable by
alternating timed automata with one clock. More interestingly, the
converse is also true: there are languages recognizable by alternating
timed automata with one clock that are not recognizable by
nondeterministic timed automata with any number of clocks.

Once the decidability of the emptiness problem for alternating timed
automata with one clock is shown, the next natural question is the
complexity of the problem. We show a non-primitive recursive lower
bound. For this we give a reduction of the reachability problem for
lossy channel systems~\cite{Sch02}. The reduction shows that the lower
bound holds also for purely universal alternating timed automata. This
implies non-primitive recursive lower bound for the universality
problem for nondeterministic timed automata with one clock. We also
point out that allowing $\epsilon$-transitions in our model permits to
code perfect channel systems and hence makes the emptiness problem
undecidable.

All this applies to automata over finite timed words.
In the case of infinite words, we prove undecidability of
the universality problem of nondeterministic automata with
one clock, by the reduction of the halting problem.
This immediately implies undecidability of the emptiness
problem for alternating one-clock automata.

\paragraph{Related work}
Our work is strongly inspired by the results of Ouaknine and
Worrell~\cite{OW04}. Techniques similar to our decidability proof and
to insights of~\cite{OW04} have been developed eariler in~\cite 
{AJ98,AN01}.

Except for~\cite{DW99}, it seems that the notion of
alternation in the context of timed automata was not studied before.
The reason was probably
undecidability of the universality problem.
The alternating automata introduced in~\cite{DW99} run over
infinite timed trees and
were used to show decidability of model checking for TCTL.
Emptiness for these automata is apparently undecidable, even
under one-clock restriction, in view of our result for one-clock
automata over infinite words.
On the other hand, emptiness for nondeterministic timed tree automata is
decidable~\cite{LN01}.

Some research
(see~\cite{AMPS98,CHR02,BDMP03,ABM04,BCFL04} and references within)
was devoted to the control problem in the timed case. While in this
case one also needs to deal with some universal branching, these works
do not seem to have direct connection to our setting.

Furthermore, let us mention that restrictions to one clock
(and two clocks) have been already considered in
the context of TCTL model-checking of timed systems~\cite{Dim00,LMS04},
leading to a lower complexity in some cases.
Finally, in~\cite{AHV93} the parametric variant of emptiness problem
was shown decidable under restriction to one clock (similarly as in our
setting) and undecidable for three clocks; the two-clock case is
left as an open question.

Similar results to ours were obtained independently by Ouaknine and
Worrell~\cite{OW05} and by Abdulla et al~\cite{ADOW05}.
The former paper defines alternating timed automata, in a slightly
different way than ours, and applies these automata to prove
decidability of model-checking for Metric Temporal Logic.
The non-primitive recursive lower bound is also established.
In the latter paper, the undecidability result for the universality
problem over infinite words is proved.

\paragraph{Organization of the paper} In the next section we
define alternating timed automata; we discuss their basic properties
and relations with nondeterministic timed automata. In
Section~\ref{sec:dec} we show decidability of the emptiness problem
for alternating timed automata with one clock. In the following
two sections we show a non-primitive recursive lower bound for the
problem,
and then the undecidability result for an extension of our model
with $\epsilon$-moves.
In Section~\ref{s:infin} we investigate automata over infinite words.

A preliminary version of this article appeared as~\cite{LW05}.

\section{Alternating Timed Automata}
\label{s:ata}

In this section we introduce the alternating timed automata model and
study its basic properties. The model is a quite straightforward
extension of the nondeterministic model. Nevertheless some care is
needed to have the desirable feature that complementation corresponds
to exchanging existential and universal branchings (and final and
non-final states). As can be expected, alternating timed automata can
recognize more languages than their nondeterministic counterparts. The
price to pay for this is that the emptiness problem becomes
undecidable, in contrast to timed automata~\cite{AluDil94}.
This motivates the restriction to automata with one
clock. With one clock alternating automata can still recognize
languages not recognizable by nondeterministic automata and moreover,
as we show in the next section, they have decidable emptiness
problem.

For a given finite set $\clo$ of {\em clock variables} (or {\em
    clocks} in short), consider the set $\constr$ of clock constraints
$\sigma$ defined by
\[
\sigma \ \ \ ::= \ \ \
\x < c \ \ | \ \
\x \leq c \ \ | \ \
\sigma_1 \land \sigma_2 \ \ | \ \ \neg \sigma,
\]
where
$c$ stands for an arbitrary nonnegative integer constant,
and $\x \in \clo$.
For instance, note that $\ttt$ (always true), or $\x = c$,
can be defined as abbreviations.
Each constraint $\sigma$ denotes a subset $\deno{\sigma}$ of $(\rplus)
^{\clo}$,
in a natural way, where $\rplus$ stands for the set of nonnegative
reals.

Transition relation of a timed automaton~\cite{AluDil94} is usually
defined
by a finite set of rules $\delta$ of the form
\[
\delta \subseteq Q \times \Sigma \times \constr \times Q \times
\powers{\clo},
\]
where $Q$ is a set of {\em locations} (control states) and $\Sigma$ is
an input alphabet.  A rule $\langle q, a, \sigma, q', r \rangle \in
\delta$ means, roughly, that when in a location $q$, if the next input
letter is $a$ and the constraint $\sigma$ is satisfied by the current
valuation of clock variables, the next location can be $q'$ and the
clocks in $r$ should be reset to $0$.  Our definition below uses an
easy observation, that the relation $\delta$ can be suitably
rearranged into a finite partial function
\[
Q \times \Sigma \times \constr \stackrel{\cdot}{\to} \powers{Q \times
\powers{\clo}}.
\]
The definition below comes naturally when one thinks of an element
of the codomain as a disjunction of a finite number of pairs
$\pair{q}{r}$.
Let $\boo{X}$ denote the set of all positive boolean formulas
over the set $X$ of propositions, i.e., the set generated by:
\[
\phi \ \ \ ::= \ \ \ X \ \ | \ \ \phi_1 \land \phi_2 \ \ | \ \
\phi_1 \lor \phi_2.
\]
%
\begin{definition}[Alternating timed automaton]
\label{d_alt}
An {\em alternating timed automaton} is a tuple $\autom = (Q, q_0,
\Sigma, \clo, F, \delta)$ where: $Q$ is a finite set of locations,
$\Sigma$ is a finite input alphabet, $\clo$ is a finite set of clock
variables, and $\d: Q\times\S\times\constr \stackrel{\cdot}{\to}
\boo{Q \times \powers{\clo}}$ is a finite partial function. Moreover
$q_0 \in Q$ is an {\em initial state} and $F \subseteq Q$ is a set of
{\em accepting states}. We also put an additional restriction:
\begin{description}
\item[(Partition)] For every $q$ and $a$, the set $\{\deno{\sigma} :
    \d(q,a,\s) \text{ is defined}\}$ gives a (finite) partition of
    $(\rplus)^\clo$.
\end{description}
\end{definition}
The (Partition) condition does not limit the expressive power of
automata. We impose it because it permits to give a nice symmetric
semantic for the automata as explained below. We will often write
rules of the automaton in a form: $\tr{q}{a}{\sigma}{b}$.

By a {\em timed word} over $\Sigma$ we mean a finite sequence
\begin{equation}
\label{e_tw}
w = \pair{a_1}{t_1} \pair{a_2}{t_2} \ldots \pair{a_n}{t_n}
\end{equation}
of pairs from $\Sigma \times \rplus$.  Each $t_i$ describes the
amount of
time that passed between reading $a_{i{-}1}$ and $a_i$, i.e., $a_1$
was read at time $t_1$, $a_2$ was read at time $t_1 {+} t_2$, and so
on. In Sections~\ref{sec:lower} and~\ref{sec:epsilon} it will be more
convenient to use an alternative representation where $t_i$ denotes
the time elapsed since the beginning of the word.
In this paper we deal with finite timed words, except
Section~\ref{s:infin}, where we will investigate timed $\omega$-words.

To define an execution of an automaton, we will need two operations
on valuations $\val \in (\rplus)^{\clo}$.
A valuation $\incr{\val}{t}$, for $t \in \rplus$, is obtained from $
\val$ by
augmenting value of each clock by $t$.
A valuation $\reset{\val}{r}$, for $r \subseteq \clo$, is obtained
by reseting values of all clocks in $r$ to zero.

For an alternating timed automaton $\autom$ and a timed word $w$ as in
(\ref{e_tw}), we define the \emph{acceptance game $\game$} between two
players \adam\ and \ewa.  Intuitively, the objective of \ewa\ is to
accept $w$, while the aim of \adam\ is the opposite.  A play starts at
the initial configuration $\conf{q_0}{\zero}$, where $\zero : \clo \to
\rplus$ is a valuation assigning $0$ to each clock variable.  It
consists of $n$ phases.  The $(k{+}1)$-th phase starts in
$\conf{q_k}{\val_k}$, ends in some configuration
$\conf{q_{k{+}1}}{\val_{k{+}1}}$ and proceeds as follows.  Let
$\bar{\val} := \val_k + t_{k{+}1}$.  Let $\s$ be the unique constraint
such that $\bar\val$ satisfies $\s$ and $b=\d(q_k,a_{k+1},\s)$ is
defined.
Existence and uniqueness of such $\s$ is implied by the (Partition)
condition.
Now the outcome of the phase is determined by
the formula $b$. There are three cases:
\begin{itemize}
\item $b = b_1 \land b_2$: \adam\ chooses one of subformulas $b_1$,
$b_2$ and the
play continues with $b$ replaced by the chosen subformula;
\item
$b = b_1 \lor b_2$: dually, \ewa\ chooses one of subformulas;
\item
$b = (q, r) \in Q \times \powers{\clo}$: the phase ends
with the result
$\conf{q_{k{+}1}}{\val_{k{+}1}} := \conf{q}{\reset{\bar{\val}}{r}}$.
A new phase is starting from this configuration if
$k{+}1 < n$.
\end{itemize}
The winner is \ewa\ if $q_n$ is accepting ($q_n \in F$), otherwise
\adam\ wins.

Formally, a play is a finite sequence of consecutive game positions  
of the form
$\langle k, q, \val \rangle$ or $\langle k, q, b \rangle$, where $k$ is
the phase number, $b$ a boolean formula, $q$ a location and $\val$ a  
valuation.
A \emph{strategy} of Eve is a mapping which assigns to each such
sequence ending in Eve's position a next move of Eve. 
A strategy is winning if Eve wins whenever she applies this strategy.

\begin{definition}[Acceptance]
The automaton $\autom$ {\em accepts} $w$ iff \ewa\ has a winning
strategy in the game $\game$.
By $\lang$ we denote the language of all timed words $w$ accepted by
$\autom$.
\end{definition}

To show the power of alternation we give an example of an automaton
for a language not recognizable by standard
(i.e. nondeterministic) timed automata (cf.~\cite{AluDil94}).

\begin{example}\label{ex:one}
    Consider a language consisting of timed words $w$ over a singleton
    alphabet $\{a\}$ that contain no pair of letters such that one of
    them is precisely one time unit later than the other.  The
    alternating automaton for this language has three states $q_0, q_1,
    q_2$. State $q_0$ is initial.  The automaton has a single clock $\x$
    and the following transition rules:

\begin{minipage}{0.47\linewidth}
\begin{eqnarray*}
\treq{q_0}{a}{\ttt}{\pair{q_0}{\emptyset} \land \pair{q_1}{\{\x\}}} \\
\treq{q_1}{a}{\x {=} 1}{\pair{q_2}{\emptyset}}
\end{eqnarray*}
\end{minipage}
\begin{minipage}{0.47\linewidth}
\begin{eqnarray*}
\treq{q_1}{a}{\x {\neq} 1}{\pair{q_1}{\emptyset}} \\
\treq{q_2}{a}{\ttt}{\pair{q_2}{\emptyset}}
\end{eqnarray*}
\end{minipage}\\

\noindent
States $q_0$ and $q_1$ are accepting, $q_2$ is not.
In state $q_0$, at each input letter, \adam\ chooses
either to stay in $q_0$ either to to go to $q_1$;
In the latter case clock $\x$ is reset.
Furthermore, the automaton can  only quit state
$q_1$ exactly one time unit after entering it.
Hence, \adam\ has a strategy to reach $q_2$ iff the word is not in
the language, i.e.,
some letter is one time unit after some other.
\end{example}
As one expects, we have the following:
\begin{proposition}
\label{p_clos}
The class of languages accepted by alternating timed automata is
effectively closed under all boolean operations: union, intersection
and complementation. These operations do not increase the number of
clocks of the automaton.
\end{proposition}

The closure under conjunction and disjunction is straightforward since
we permit positive boolean expressions as values of the transition
function. Due to the condition (Partition) the automaton $\neg \autom$
for the
complement is obtained from $\autom$ by exchanging conjunctions with
disjunctions in
all transitions and exchanging accepting states with non-accepting
states.

\begin{definition}
An alternating timed automaton $\autom$ is called {\em purely universal}
if the disjunction does not appear in the transition rules $\delta$.
Dually, $\autom$ is {\em purely existential} if no conjunction appears
in $\delta$.
\end{definition}

Clearly, if $\autom$ is purely universal (purely existential) then $\neg
\autom$ is purely existential (purely universal).
It is obvious that every purely existential automaton is a standard
nondeterministic timed automaton. The converse requires a proof
because of the (Partition) condition.

\begin{proposition}
    Every standard nondeterministic automaton is equivalent to a
    purely existential automaton.
\end{proposition}
\begin{proof}
Transition relation of a nondeterministic timed automaton is usually
defined
by a finite set $\delta$ of rules of the form
$\langle q, a, \sigma, q', r \rangle \in
Q \times \Sigma \times \constr \times Q \times \powers{\clo}.
$
Given such an automaton $\autom$, the corresponding purely
existential alternating
automaton $\widehat{\autom}$ has the same set $Q$ of states
as $\autom$, plus one additional state
$q_{\mathtt{sink}}$.
Automaton $\widehat{\autom}$ has the same initial state and accepting
states as $\autom$,
the same set of clocks $\clo$, and the same input alphabet.
The only essential difference is that $\d$ is replaced by
$\widehat{\d}: Q\times\S\times\constr \stackrel{\cdot}{\to}
\boo{Q \times \powers{\clo}}$, defined as follows.

In fact, we prefer to define $\widehat{\d}$ equivalently as
$\widehat{\d}: Q\times\S\times\constr \stackrel{\cdot}{\to}
\powers{Q \times \powers{\clo}}$.
\removed{
For convenience assume that for each pair $(q, a)$,
the constraints applicable in $\autom$ at $(q, a)$ cover the whole set
$(\rplus)^\clo$ of clock valuations, i.e.,
\[
\{ \deno{\s} : \langle q, a, \s, q', r \rangle \in \d,
\text{\ \ for some \ } q', r  \} = (\rplus)^\clo.
\]
Automaton $\autom$ can be easily ...
}
Let $\s_1 \ldots \s_n$ be all clock constraints appearing in
$\d$.
The guards appearing in $\widehat{\d}$ will be $\s_X$, for $X
\subseteq \{1 \ldots n\}$,
defined by:
\[
\s_X = \land_{i \in X} \s_i \ \ \land \ \ \land_{i \notin X} \neg \s_i.
\]
I.e., we consider conjunctions of arbitrary sets of guards $\s_i$.
The value $\widehat{\d}(q, a, \s)$ is defined iff $\s = \s_X$ for
some $X$,
hence $\widehat{\d}$ clearly satisfies the (Partition) condition.
The constraints $\s_X$ satisfying $\deno{\s_X} = \emptyset$ can be
safely omitted.
We put:
\[
\widehat{\d}(q, a, \s_X) = \{ (q', r) :
\langle q, a, \s_i, q', r \rangle \in \d \text{\ \ for some \ } i \in
X  \}.
\]
If $\widehat{\d}(q, a, \s_X)$ is empty, we put
$\widehat{\d}(q, a, \s_X) = \{ (q_{\mathtt{sink}}, \emptyset) \}$.
And finally we put: $\widehat{\d}(q_{\mathtt{sink}}, a, \s_X) =
\{ (q_{\mathtt{sink}}, \emptyset) \}$, for any $a$ and $\s_X$.

It is routine now to check that languages accepted by
$\autom$ and $\widehat{\autom}$ coincide.
\qed
\end{proof}

In the following sections, we consider emptiness, universality and
containment for different classes of alternating timed automata.
For clarity, we recall definitions here.
\begin{definition}
For a class $C$ of automata we consider three problems:
\begin{itemize}
\item Emptiness: given $\Aa\in C$ is $L(\Aa)$ empty.

\item Universality: given $\Aa\in C$ does $L(\Aa)$ contain all timed
    words.

\item Containment: given $\Aa,\Bb\in C$ does $L(\Aa)\subseteq L(\Bb)$.
\end{itemize}
\end{definition}
It is well known that the universality is undecidable for
non-deterministic timed automata~\cite{AluDil94} with at
least two clocks. As a consequence, all three problems are
undecidable for alternating timed automata with two clocks. This is
why, in the rest of the paper, we focus on automata with one clock only.

\paragraph{Proviso:} In the following all automata have one clock.\\

\noindent
The automaton from
Example~\ref{ex:one} uses only one clock. This shows that one clock
alternating automata can recognize some languages not recognizable by
nondeterministic automata with many clocks. The converse is also true:
\begin{theorem}
Classes of languages recognizable by nondeterministic timed automata
and by one-clock alternating timed automata are incomparable.
\end{theorem}
%
%
\begin{proof}
We show a language acceptable by a deterministic automaton with
many clocks but not acceptable by an alternating automaton with one
clock.

Consider the timed language over the singleton alphabet $\{b \}$
consisting of the words containing appearances of the letter $b$ at
times $t_1$ and $t_2$, where $0<t_1< t_2<1$, no other $b$ in between
$0$ and $1$ and precisely one $b$ between $t_1+1$ and $t_2+1$. We will
show that this language cannot be accepted by an alternating timed
automaton with one clock. Obviously it is accepted by a
deterministic timed automaton with two clocks.

For a preparation consider a deterministic untimed automaton $\Bb$.
A sequence $b^k$ of $k$ letters $b$ determines a function
$f^{\Bb}_k:Q^\Bb\to Q^\Bb$ saying that if started in the state $q$ after
reading $b^k$ the automaton will end in $f^{\Bb}_k(q)$. Clearly the
number of
such functions is bounded if the number of states is fixed. Thus there
are $m$ and $l$, depending only on the number of states, such that
$f^\Bb_m=f^\Bb_{m+l}$.  Moreover $f^\Bb_{m+i}=f^\Bb_{m+l+i}$ for all
$i>0$.

To arrive at a contradiction assume that our language is recognized
by an ATA $\Aa$ with $n$ states. Suppose for a moment that all
constants in the tests in transition function of the automaton are
integers. Let $m$ and $l$ be such that $f^\Bb_{m+i}=f^\Bb_{m+l+i}$ for
all $i>0$ and for all deterministic automata $\Bb$ with at most
$2^{2^{2n}}$ states.

Now consider two words $w_1$ and $w_2$. In $w_1$ we have $b$ at times
$0.3$, $0.7$, $1.5$ and $m$ $b$'s somewhere in the interval $(1,1.3)$
as well as $m$ $b$'s somewhere in the interval $(1.7,2)$. Word $w_2$
is obtained from $w_1$ by adding $l$ $b$'s somewhere in the interval
$(1.3,1.7)$; but not at point $1.5$ of course. We will show that if
$\Aa$ accepts $w_1$ then it also accepts $w_2$.

Consider the accepting run of $\Aa$ on $w_1$. Look at the
configurations in which the automaton reaches at time $1$. Let $(q,v)$
be one of them. The value of the clock $v$ can be $0.3$, $0.7$ or
$1$. This is because there are only two letters till $1$ and the
automaton can reset clock only when it reads a letter. We will analyse
the three cases one by one.

If $v=1$ then it is easy to see that from a configuration $(q,v)$ the
automaton has no use for the clock in the interval $(1,2)$. If not
reset, the value of the clock in this interval will be in $(1,2)$ and
the automaton can compare the values only with integers. If the clock
is reset then its value will stay in $(0,1)$ till the end of the
interval. Thus from the configuration $(q,v)$ automaton $\Aa$ behaves
as an alternating automaton without a clock with additional flag
telling whether there was a reset or not. Because it has $n$ states,
it is equivalent to a deterministic automaton of at most $2^{2^{2n}}$
states. We have that if it accepts from $q$ the string of $2m+1$
letters $b$ then it also accepts $2m+l+1$ letters $b$. Thus $\Aa$ has
an accepting run from $(q,v)$ in $w_2$ if it had one in $w_1$.

If $v=0.7$ then consider the run of $\Aa$ from $(q,v)$ till the time
point $1.3$. Automaton $\Aa$ has no use of the clock till that point
for the same reason as above.  It arrives at a set of configurations:
some with the value of the clock $1$ and some with the value
$<0.3$. The later are possible because $\Aa$ could reset a
clock. Consider the rest of the computation starting from a
configuration $(q',1)$. Once again the clock will not be useful to
$\Aa$ in the rest of the word. Hence we will arrive to the same final
states on $a^{1+m}$ and $a^{1+m+l}$. Similarly for all the
configurations with the values of the clock $<.3$.

If $v=0.3$ then consider the run of $\Aa$ from $(q,v)$ till the time
point $1.7$. Till that time there was no use of the clock. We get a
set of configurations with clock value $1$ and the other with clock
value $<0.7$.
The possible configurations with clock value $1$ are the same
no matter if we have made automaton run on $w_1$ or on $w_2$,
for the same reason as before.
As the rest of $w_1$ is the same as the rest of $w_2$ we are done.
On the other hand, when comparing configurations with clock value
$<0.7$ in runs over $w_1$ and $w_2$, the possible locations are
the same but the clock values may differ. But the clock value is
irrelevant
before time $2$, hence again we are done.
%
%

In the argument we essentially use the assumption that
we compare clocks only with natural numbers. If we allowed to
compare with rationals we can get an example of the similar kind by
using rescaling. Instead of intervals $(0,1)$ and $(1,2)$ we would use
smaller intervals that are of the size smaller than the smallest
constant used by the automaton.

More precisely, let $c \neq 0$ be the smallest positive rational
such that the clock is compared in $\Aa$ either to
$c$ or to $1 {-} c$ or to $1 {+} c$.
We define words $w_1$ and $w_2$ as follows.
In $w_1$ we have $b$ at times
$0.3 c$, $0.7 c$, $1 + 0.5 c$ and $m$ $b$'s somewhere in the interval
$(1,1 + 0.3 c)$
as well as $m$ $b$'s somewhere in the interval $(1 + 0.7 c, 1 + c)$.
Word $w_2$
is obtained from $w_1$ by adding $l$ $b$'s somewhere in the interval
$(1 + 0.3 c,1 + 0.7 c)$; but not at point $1 + 0.5 c$.
The whole proof works unchanged.
\qed
\end{proof}

%
%

\section{Decidability}\label{sec:dec}

The main result of this section is that the emptiness problem for
one-clock alternating timed automata is decidable. Due to closure
under boolean operations, this implies
the decidability of the
universality and the containment problems.

\begin{theorem}
\label{t_decid}
The emptiness problem is decidable for one-clock alternating timed
automata.
\end{theorem}

\begin{corollary}
    The containment problem is decidable for one-clock alternating timed
    automata.
\end{corollary}

The rest of this section is devoted to the proof of
Theorem~\ref{t_decid}.  Essentially, we have adapted the method of
Ouaknine and Worrell~\cite{OW04} for our more general setting. We
point out the differences below.

Fix a one-clock alternating timed automaton $\autom = (Q, q_0, \Sigma,
\{\x\}, F, \delta)$.  For readability, assume w.l.o.g.\ that the
boolean conditions appearing in rules of $\delta$ are all in {\em
    disjunctive normal form}.  In terms of acceptance games this means
that each phase consists of a single move of \ewa\ followed by a
single move of \adam.  Consider a labelled transition system $\Tt$
whose states are finite sets of configurations, i.e., finite sets of
pairs $\conf{q}{\val}$, where $q \in Q$ and $\val \in \rplus$.  The
initial position in $\Tt$ is $P_0 = \{ \conf{q_0}{\mathbf{0}} \}$ and
there is
a transition $\tran{P}{\pairnb{a}{t}}{P'}$ in $\Tt$ iff $P'$ can be
obtained from $P$ by the following nondeterministic process:
\begin{itemize}
\item
First, for each $\conf{q}{\val} \in P$, do the following:
\begin{itemize}
\item
let $\val' := \incr{\val}{t}$,
\item
let $b=\d(q,a,\s)$ for the uniquely determined $\s$ satisfied in
$\val'$,
\item
choose one of disjuncts of $b$, say
\[
\pair{q_1}{r_1} \ \land \ \ldots \ \land \ \pair{q_k}{r_k}
\ \ \ \ \ (k > 0),
\]
\item
let $\next{} = \{ \conf{q_i}{\reset{\val'}{r_i}} : i = 1 \ldots k \}$.
\end{itemize}
\item
Then, let $P' := \bigcup_{\conf{q}{\val} \in P} \next{}$.
\end{itemize}

This construction is very similar to the translation from alternating
to nondeterministic automata over (untimed) words: we just
collect all universal choices in one set.  Compared to~\cite{OW04},
the essential difference is that we have to deal with both disjunction
and conjunction, while in~\cite{OW04} only one of them appeared.  We
treat conjunction similarly to determinization in~\cite{OW04}.
On the other hand, we leave the existential choice, i.e.,
nondeterminism, essentially unaffected in $\Tt$.

In what follows we will derive from $\Tt$ a finite-branching
transition system $\HH$, suitable for the decision procedure.
Like in~\cite{OW04}, the degree of the nodes of $\HH$ will not
be bounded but
nevertheless finite. This is sufficient for our purposes.

A state $\{ \conf{q_1}{\val_1}, \ldots, \conf{q_n}{\val_n}\}$ of $\Tt$
is called {\em bad} iff all control states $q_i$ are accepting ($q_i
\in F$). The following proposition characterizes acceptance in $\Aa$
in terms of reachability of bad states in $\Tt$.
It is enough to consider reachability because $\Aa$ accepts only
finite words.
\begin{lemma}
\label{lem:reach}
    $\Aa$ accepts a timed word $w$ iff there is a path in $\Tt$,  
labelled
    by $w$, from $P_0$ to a bad state.
\end{lemma}
Let $\TT$ be a labelled transition system obtained from $\Tt$ by
erasing time information from transition labels, i.e., there is a
transition
$\tran{P}{a}{Q}$ in $\TT$ iff there is $\tran{P}{\pairnb{a}{t}}{Q}$
in $\Tt$,
for some $t \in \rplus$.  Now we cannot talk about particular timed
words but still we have the following:
\begin{lemma}\label{lemma:p_iff}
    $\lang$ is nonempty if and only if there is a path in $\TT$ from
    $P_0$ to a bad state.
\end{lemma}
Thus, the (non)emptiness problem for $\Aa$ is reduced to the
reachability of a bad state in $\TT$.  The last difficulty is that
even if each state of $\TT$ is a finite set, there are uncountably
many states. The following definition allows to abstract from the
precise timing information in states.

Let $\cmax$ denote the biggest
constant appearing in constraints in $\delta$. Let set $\reg$ of
{\em regions} be a partition of $\rplus$ into $2 \cdot (\cmax{+}1)$
sets as
follows:
\[
\reg := \{ \{0\}, (0,1), \{1\}, (1,2), \ldots,
(\cmax{-}1, \cmax), \{\cmax\}, (\cmax, +\infty) \}.
\]
For $\val \in \rplus$, let $\reg(\val)$ denote its region;
and let $\fract{\val}$ denote the fractional part of $\val$.
Below we work with finite words over the alphabet $\alp=\powers{Q
\times \reg}$ consisting of finite sets of pairs $\pair{q}{\rreg}$,
where $q \in Q$
is a control state and $\rreg \in \reg$ is a region.
\begin{definition}
\label{d:H}
For a state $P$ of $\TT$ we define a word $H(P)$ from $\alp^*$ as
the one obtained by the following procedure:
\begin{itemize}
\item
replace each $\pair{q}{\val} \in P$ by a triple
$\langle q, \reg(\val), \fract{\val} \rangle$
(this yields a finite set of triples)
\item
sort all these triples w.r.t.\ $\fract{\val}$
(this yields a finite sequence of triples)
\item
group together triples that have the same value of $\fract{\val}$,
ignoring multiple occurrences
(this yields a finite sequence of finite sets of triples)
\item
forget about $\fract{\val}$, i.e., replace each triple
$\langle q, \reg(\val), \fract{\val} \rangle$ by a
pair $\pair{q}{\reg(\val)}$
(this yields a finite sequence of finite sets of pairs,
a word in $\alp^*$).
\end{itemize}
\end{definition}
\begin{example}
To illustrate transformation $H$, consider
$P = \{ (q_1, 0.5), (q_2, 1.2),$ $(q_3, 2.2) \}$,
where $q_1, q_2, q_3$ are locations.

Let $\cmax = 2$. Denote regions by $\creg{0} = \{0\}$,
$\creg{0,1} = (0,1), \ldots, \creg{2} = \{2\}, \creg{2,+\infty} = (2,  
+\infty)$.
First, $P$ is transformed into the set
$$\{ \langle q_1, \creg{0,1}, 0.5 \rangle, \langle q_2, \creg{1,2},  
0.2 \rangle,
\langle q_3, \creg{2,+\infty}, 0.2 \rangle \}.$$
\noindent
We make it into a sorted sequence
$\langle q_2, \creg{1,2}, 0.2 \rangle
\langle q_3, \creg{2,+\infty}, 0.2 \rangle \langle q_1, \creg{0,1},  
0.5 \rangle$.
Then we group together triples with the same fractional part,
obtraining a sequence of length two:
\[
\{ \langle q_2, \creg{1,2}, 0.2 \rangle,
\langle q_3, \creg{2,+\infty}, 0.2 \rangle \}, \{ \langle q_1, \creg 
{0,1}, 0.5 \rangle \}.
\]
Finally we remove the fractional parts and obtain
\[
H(P) = \{ (q_2, \creg{1,2}), (q_3, \creg{2,+\infty}) \}, \{ (q_1,  
\creg{0,1}) \}.
\]
\end{example}
\begin{definition}
Let $\HH$ be the transition system whose states are words $H(P)$ for
$P$ a state of $\TT$; a transition $\tran{W_1}{a}{W_2}$ is in
$\HH$ if there is a transition $\tran{P_1}{a}{P_2}$ in $\TT$ with
$H(P_1) = W_1$, $H(P_2) = W_2$. The initial state in $\HH$ is $W_0 = H 
(P_0)$.
\end{definition}
\begin{example}
Assume that the automation from previous example has a rule:
\[
\tr{q_3}{a}{\x {>} 2}{\pair{q_1}{\x} \,\lor\, (\pair{q_2}{\emptyset}  
\land \pair{q_3}{\emptyset})}.
\]
Imagine a transition $\tran{P}{a}{P'}$ in $\TT$ corresponding to
$\tran{P}{\pairnb{a}{0.6}}{P'}$ in $\Tt$ derived from the above rule.
There are two possibilities:
$P' = \{ (q_1, 1.1), (q_2, 1.8), (q_1, 0) \}$ or
$P' = \{ (q_1, 1.1), (q_2, 1.8),$ $(q_2, 2.8), (q_3, 2.8) \}$.
Accordingly, there are two transitions $\tran{H(P)}{a}{W'}$ in $\HH$,  
for
$W' = \{ (q_1, \creg{0}) \} \{ (q_1, \creg{1,2}) \} \{ (q_2, \creg 
{1,2}) \}$
or
$W' = \{ (q_1, \creg{1,2}) \} \{ (q_2, \creg{1,2}), (q_2, \creg{2,+ 
\infty}), (q_3, \creg{2,+\infty}) \}$.
In each case $W' = H(P')$. Hence, transitions in $\HH$ can
``simulate'' transitions in $\TT$.
On the other hand, $H(P)$ has also a transition
\[
\tran{H(P)}{}{\{ (q_1, \creg{0}) \} \{ (q_1, \creg{1,2}) \}
\{ (q_2, \creg{1,2}), (q_2, \creg{2,+\infty}), (q_3, \creg{2,+ 
\infty}) \}}
\]
that simulates a posible transition of
$\bar{P} =  \{ (q_1, 0.5), (q_2, 1.2), (q_3, 2.2), (q_3, 6.2) \}$.
Hence, roughly speaking, transitions of $H(P)$ correspond to the union
of all the transitions of all $\bar{P}$ such that $H(\bar{P}) = H(P)$.
\end{example}
If $P$ is bad and $H(P) = H(P')$ then $P'$ is bad as well.
Hence it is correct to call a state $W$ in
$\HH$ {\em bad} if $W = H(P)$ for a bad state $P$.

\removed{

\begin{lemma}\label{lemma:zamiast}
A bad state is reachable in $\TT$ from $P_0$ iff
a bad state is reachable in $\HH$ from $W_0$.
\end{lemma}
\begin{definition}
Define an equivalence relation $\equ$ over states of $\TT$ as
the kernel of $H$, i.e., $P \equ P'$ iff $H(P) = H(P')$.
\end{definition}
The following observations are straightforward:
\begin{lemma}
\label{lem:bisim}   Relation $\equ$ is a bisimulation over transition  
system $\TT$.
\end{lemma}

\begin{lemma}\label{lemma:presbad}
    If $P$ is bad and $P\equ P'$ then $P'$ is bad.
\end{lemma}
Let $\HH$ denote the quotient of the transition system $\TT$ by $\equ 
$. To
put it more explicitly: states of $\HH$ are all words $H(P)$, for a
state $P$ of $\TT$; there is a transition $\tran{W_1}{a}{W_2}$ in
$\HH$ if there is a transition $\tran{P_1}{a}{P_2}$ in $\TT$ with
$H(P_1) = W_1$, $H(P_2) = W_2$. Since $\equ$ is a bisimulation, the
definition does not depend on a particular choice of $P_1$ (and $P_2$).
The initial state $W_0$ in $\HH$ is $H(P_0)$.

By Lemma~\ref{lemma:presbad} it is correct to call a state $W$ in
$\HH$ {\em bad} if $W = H(P)$ for a bad state $P$.  Because $\HH$
is a quotient of $\TT$ by bisimulation, from
By Lemma~\ref{lemma:zamiast} we derive:

} 

\begin{lemma}\label{lemma:iff2}
$\lang$ is nonempty iff
a bad state is reachable
in $\HH$ from $W_0$.
\end{lemma}
\begin{proof}
By Lemma~\ref{lemma:p_iff} we only need to show:
a bad state is reachable in $\TT$ from $P_0$ iff
a bad state is reachable in $\HH$ from $W_0$.

Consider a transition system $\Tt'$ obtained from $\Tt$ by imposing
one additional restriction on transitions:
whenever $\val_1$ and $\val_2$ are in the same region, then $\next{1}  
= \next{2}$.
By $\TT'$ and $\HH'$ denote the transition systems obtained from $ 
\Tt'$ instead
of $\Tt$.
They have the same states as $\TT$ and $\HH$, respectively, but fewer  
transitions.
Clearly, the additional restriction has no impact on acceptance,  
i.e., on reachability of a bad state.
Hence we have:
a bad state is reachable in $\TT$ from $P_0$ iff a bad state is  
reachable in $\TT'$ from $P_0$.
And also: a bad state is reachable in $\HH$ from $W_0$ iff a bad  
state is reachable in $\HH'$ from $W_0$.

Now observe that the graph of $H$, i.e., the set of all pairs $(P, H 
(P))$, is a bisimulation
between $\TT'$ and $\HH'$.
If $\tran{P}{a}{P'}$ then obviously $\tran{H(P)}{a}{H(P')}$.
If $\tran{H(P)}{a}{W'}$ then there exists $P'$ such that $\tran{P}{a} 
{P'}$ and
$H(P') = W'$; we only need to guess appropriate $t$ and derive $P'$  
from transition
$\tran{P}{\pairnb{a}{t}}{P'}$ in $\Tt'$ (clearly $t$ need not be  
unique).

The bisimulation guarantees that
a bad state is reachable in $\TT'$ from $P_0$ iff
a bad state is reachable in $\HH'$ from $W_0$.
This completes the proof.
\qed
\end{proof}
At this point, we have reduced emptiness of $\lang$ to the
reachability of a bad state in a countably infinite transition system
$\HH$.  The rest of the proof is quite standard~\cite{ACJT96,FinSch01}
and exploits the fact that one can put an appropriate {\em well-quasi-
order}
({\em wqo} in short) on
states of $\HH$. Unfortunately, we are obliged to redo the proofs as
we could not find a theorem that fits precisely our setting.
\begin{definition}\label{d_wqo}
    Let $\wqo$ denote the {\em monotone domination} ordering over
    $\alp^*$ induced by the subset inclusion over $\alp$, defined as
    follows: $a_1 \ldots a_n \wqo b_1 \ldots b_m$ iff there exists a
    strictly increasing function $f : \{ 1, \ldots, n\} \to \{ 1,
    \ldots, m \}$ such that for each $i \leq n$, $a_i \subseteq
    b_{f(i)}$.
\end{definition}
\begin{lemma}[\cite{Hig52}]\label{l_Hig}
    Relation $\wqo$ is a wqo, i.e., for arbitrary infinite sequence
    $W_1, W_2, \ldots$ of words over $\alp$, there exist indexes $i < j$
    such that $W_i \wqo W_j$.
\end{lemma}
The decision procedure for reachability of bad states will work
by an exhaustive search through a sufficiently large portion of the
whole reachability tree.  Thus we need to know that an arbitrarily
large part of that tree can be effectively constructed.
Roughly, all time delays of an action $a$ from $W$
can be captured by a finite number of cyclic shifts of $W$
with an appropriate change of region.
\begin{lemma}
\label{lemma:succ}
    For each state $W$ in $\HH$, its set of successors $\{ W' \in \alp^*
    : \tran{W}{a}{W'} \text{\ \ for some \ } a \}$ is finite and
    effectively computable.
\end{lemma}
\begin{proof}
Recall that a word $W$ represents a finite set of pairs $(q, \val)$.
The letters are sorted according to the value of $\fract{\val}$;
moreover the letters represent finite sets of pairs in fact,
i.e., all the pairs with the same $\fract{\val}$.
Note that all pairs with $\fract{\val} = 0$, if any, are represented by
the first letter of $W$; and the corresponding region
is of the form $\{i\}$ (or $(\cmax, \infty)$) in this case.

Now imagine a transition $\tran{W}{a}{W'}$ in $\HH$.
This corresponds to some transition $\tran{P}{a,t}{P'}$ in $\Tt$,
for some $t$ and some chosen set $P$ of pairs $(q, \val)$.
Importantly, the same time delay $t$ is applied to all the pairs $(q,
\val)$.
Denote by $\widehat{P}$ the set obtained from $P$ by time delay, i.e.,
by replacing each $(q, \val)$ with $(q, \val + t)$;
consider this, conceptually, for all $t > 0$.
The corresponding word $\widehat{W}$ in $\HH$
is obtained from $W$ by an operation similar to a cyclic shift, to
the right,
repeated as many times as needed.
This operation modifies $W$ as follows.
Note that the first letter of $W$ contains either only
pairs of the form $(q, \{i\})$, either only the pairs of the form
$(q, (i, i+1))$ (and perhaps $(\cmax, \infty)$ as well).
In the first case, change each region $\{i\}$
in the first letter of $W$ to $(i, i+1)$
(or to $(\cmax, \infty)$, if $i = \cmax$).
In the second case,
remove the right-most letter and put it as the first letter in the word,
and change each region $(i,i+1)$ to $\{i+1\}$.

Hence, the set $\{ W' \in \alp^* : \tran{W}{a}{W'} \text{\ \ for some
\ } a \}$
can be computed by applying the operation defined above an arbitrary
number of times (until all regions are $(\cmax, \infty)$), yielding $
\widehat{W}$;
and by calculating the effect of performing any transition $a$ from
$\widehat{W}$.
\qed
\end{proof}

The following observation is proved in the same way as Lemma~15
in~\cite{OW04}.
\begin{lemma}\label{lemma:sim}
    The inverse of $\wqo$ relation is a simulation:
whenever $W_1 \wqo W_2$ and $\tran{W_2}{a}{W'_2}$, there is some
$W'_1$ such that $\tran{W_1}{a}{W'_1}$ and $W'_1 \wqo W'_2$.
\end{lemma}
\begin{proof}
    Take $W_1\wqo W_2$ and suppose $\tran{W_2}{a}{W'_2}$. By definition
    it means that there is $P_2$ with $H(P_2)=W_2$ such that there is a
    transition $\tran{P_2}{a}{P'_2}$ and $H(P'_2)=W'_2$. Since $W_1\wqo
    W_2$ it is easy to see that there is $P_1\subseteq P_2$ such that
    $W=H(P_1)$; $P_1$ is obtained by removing from $P_2$ the pairs
that do not end
    up in $W_1$ when construction $H$ is applied
    (cf.~Definition~\ref{d:H}). Now, directly from the definition of the
    transition system $\TT$ we have $\tran{P_1}{a}{P'_1}$ with
    $P'_1\subseteq P'_2$. So $\tran{W_1}{a}{H(P'_1)}$. As $P'_1
\subseteq P_2$,
    we have $H(P'_1)\wqo W_2$ as required.

\removed{

We combine two simple observations.
First, $H(P_1) \wqo H(P_2)$ iff
$H(P_1) = H(P)$, for some $P \subseteq P_2$, i.e.,
$P_1 \equ P$, for some $P \subseteq P_2$.
Second, whenever $\tran{P_2}{a,t}{P'_2}$ and $P \subseteq P_2$,
there exists $P' \subseteq P'_2$ with
$\tran{P}{a,t}{P'}$.

Assume $W_1 \wqo W_2$;
we have some $P_1$, $P_2$ and $P$ as above, with
$H(P_i) = W_i$.
Assume furthermore $\tran{W_2}{a}{W'_2}$;
hence $\tran{P_2}{a,t_2}{P'_2}$, for some $t_2$ such that
$H(P'_2) = W'_2$.
It follows $\tran{P}{a,t_2}{P'}$ for some $P' \subseteq P'_2$.
And, by Lemma~\ref{lem:bisim},
$\tran{P_1}{a,t_1}{P'_1}$ for some $t_1$ and $P'_1$
such that $H(P'_1) = H(P')$.
Hence, putting $W'_1 = H(P'_1)$, we get
$\tran{W_1}{a}{W'_1}$ and $W'_1 \wqo W'_2$.

} 

\qed
\end{proof}

The next observation is more specific to our setting but fortunately
very easy.
\begin{lemma}[Downward closedness of badness]\label{lemma:down}
    Whenever $W \wqo W'$ and $W'$ is bad then $W$ is bad as well.
\end{lemma}
\begin{proof}
    Take a letter $w_i$ of $W$. We need to show that $q\in F$ for every
    $(q,\rreg)\in w_i$. By the definition of $W\wqo W'$ we have
    $w_i\subseteq w'_j$ for some letter $w'_j$ of $W'$.  Hence,
    $(q,\rreg)\in w'_j$ and $q\in F$ as $W'$ is bad.
\qed
\end{proof}

Now we are ready to prove the main lemma.
\begin{lemma}\label{l_decid}
    It is decidable whether a bad state is reachable in $\HH$ from
    $W_0$.
\end{lemma}
\begin{proof}
The {\em reachability tree} is the unravelling of $\HH$ from $W_0$.
The algorithm constructs a portion $t$ of the tree conforming to
the following rule: do not add a node $W'$ to $t$ in a situation
when among its ancestors there is some $W \wqo W'$.
Lemma~\ref{l_Hig} guarantees that each path in $t$ is finite.
Furthermore, since the degree of each node is finite, $t$ is a finite
tree.

We need only to prove that if a bad state is reachable in $\HH$ from
$W_0$ then $t$ contains at least one bad state.  Let $W$ be such a bad
state reachable from $W_0$ in $\HH$ by a path $\pi$ of the shortest
length.  Assume that $W$ is not in $t$, i.e., there are two other
nodes in $\pi$, say $W_1$ and $W_2$ such that $W_1$ is an ancestor of
$W_2$ in the reachability tree and $W_1 \wqo W_2$ (i.e., $W_2$ was {\em
    not} added into $t$).  Since the inverse of $\wqo$ is a simulation
by Lemma~\ref{lemma:sim}, the sequence of transitions in $\pi$ from
$W_2$ to $W$ can be imitated by the corresponding sequence of
transitions from $W_1$ to some other $W' \wqo W$.  $W'$ is bad as well
by Lemma~\ref{lemma:down}.  Moreover, the path leading to $W'$ is
strictly shorter than $\pi$, a contradiction.  \qed
\end{proof}
Theorem~\ref{t_decid} follows immediately from Lemma~\ref{l_decid} and
Lemma~\ref{lemma:iff2}.
\paragraph{Remark:}
In fact, Ouaknine and Worrell showed decidability of containment ''
$\lang \subseteq \langpar{\bautom}$'' in a slightly more general  
case, namely
when automaton $\autom$ has arbitrarily many clocks.  Along the same
lines one can adapt our proof,
assumed that $\autom$ is an arbitrary
nondeterministic timed automaton and $\bautom$ is a one-clock
alternating timed automaton.
We sketch below the necessary modifications.

If we denote by $\bar{\bautom}$ a dual of $\bautom$, i.e., an automaton
accepting the complement of $L(\bautom)$, then the containment reduces
to emptiness of $L(\autom) \cap L(\bar{\bautom})$.
Compared to the proof above, each state $P$ of $\Tt$ needs to contain  
additionally information
on a configuration of $\autom$.
Due to the fact that $\autom$ is purely existential,
$P$ will contain precisely one pair $(q, \val)$, where $q$ is a state
of $\autom$ and $\val$ a valuation of all its clocks.
The transition relation $\tran{P}{a,t}{P'}$ is adapted so that
the delay $t$ before performing an action $a$ is the same in $\autom$
and $\bautom$.
This guarantees that the facts analogous to Lemma~\ref{lem:reach} and~
\ref{lemma:p_iff}
hold; but now a state $P$ is {\em bad} iff all states of \emph{both} $
\autom$ and $\bautom$
appearing in $P$ are accepting.

Definition of $H$ is precisely as before, but it needs a preprocessing:
the pair $(q, \val)$ corresponding to $\autom$ is split into a number
of triples $(q, \val_x, x)$, one for each clock $x$ of $\autom$.
The triples are identical on the first component, and
$\val_x$ is the value of clock $x$.
Observe that the number of such triples is the same in each state of $ 
\HH$,
and equal to the number of clocks in $\autom$.
An analog of Lemma~\ref{lemma:iff2} holds: $L(\autom) \cap L(\bar
{\bautom})$ is nonempty
iff a bad state is reachable in $\HH$.

Finally, Lemma~\ref{lemma:succ} and~\ref{lemma:down} hold as well,
and the proofs are similar.
The proofs of Lemma~\ref{lemma:sim} and~\ref{l_decid} rest unchanged.

\section{Lower Bound}\label{sec:lower}

In this section we prove the following lower bound result.

\begin{theorem}
\label{t_lb}
The complexity of the emptiness problem for one-clock purely universal
alternating timed automata is not bounded by a primitive recursive
function.
\end{theorem}
Since emptiness and universality are dual in the setting of
alternating automata, as a direct conclusion we get the following:
\begin{corollary}
The complexity of the universality problem for one-clock purely
existential
alternating (i.e., nondeterministic) timed automata is not bounded by
a primitive recursive function.
\end{corollary}
This answers the question posed by Ouaknine and Worrell~\cite{OW04}.

The rest of this section contains the proof of Theorem~\ref{t_lb}.
The proof is a reduction of the reachability problem for \emph{lossy
    one-channel systems}~\cite{Sch02}.
\begin{definition}[Channel system]
    A {\em channel system} is given by a tuple $\syst = (Q, q_0, \Sigma,
    \Delta)$, where $Q$ is a finite set of control states, $q_0 \in Q$
    is an initial state, $\Sigma$ is a finite channel alphabet and
    $\Delta \subseteq Q \times (\{\wri{a}: a {\in} \Sigma\} \cup
    \{\rea{a}: a {\in} \Sigma\} \cup \{ \epsilon \}) \times Q$ is a
    finite set of transition rules.
\end{definition}
A configuration of $\syst$ is a pair $\pair{q}{w}$ of a control state
$q$ and a channel content $w \in \Sigma^*$.  Transition rules allow
the system to pass from one configuration to another.  In particular,
a rule $\langle q, \wri{a}, q'\rangle$ allows in a state $q$ to write
to the channel and to pass to the new state $q'$. Similarly, $\langle
q, \rea{a}, q' \rangle$ means reading from a channel and is allowed in
state $q$ only when $a$ is at the end of the channel.  The channel is
a FIFO, and by convention $\syst$ writes at the beginning and reads at
the end.  Finally, a rule $\langle q, \epsilon, q'\rangle$ allows for
a silent change of control state, without reading or writing.

Formally, there is a (perfect)
transition $\trperf{\pair{q}{w}}{\gamma}{\pair{q'}{w'}}$
if one of the following conditions is satisfied:
\begin{itemize}
\item
$\gamma = \langle q, \epsilon, q' \rangle$ and $w = w'$, or
\item
$\gamma = \langle q, \wri{a}, q' \rangle$ for some $a {\in} \Sigma$,
and $w' = aw$, or
\item
$\gamma = \langle q, \rea{a}, q' \rangle$ for some $a {\in} \Sigma$,
and $w = w'a$.
\end{itemize}

The {\em initial configuration} is $\pair{q_0}{\epsilon}$, i.e.,
execution of $\syst$ starts with the empty channel.  For technical
convenience, we assume w.l.o.g.\ that there is no rule returning back
to the initial state: for each rule $\langle q, x, q' \rangle
\in \Delta$, $q' \neq q_0$.

A {\em lossy channel system} differs from the perfect one in only one
respect: during the transition step, an arbitrary number of messages
stored in the channel may be lost.  To define lossy transitions, we
need the subsequence ordering on $\Sigma^*$, denoted by $\subse$
(e.g., $\text{\tt{tata}} \subse \text{\tt{atlanta}}$).  We say that
there is a lossy transition from $\pair{q}{w}$ to $\pair{q'}{w'}$,
denoted by $\trlossy{\pair{q}{w}}{\gamma}{\pair{q'}{w'}}$, iff there
exists $u, u' \in \Sigma^*$ such that $u \subse w$,
$\trperf{\pair{q}{u}}{\gamma}{\pair{q'}{u'}}$ and $w' \subse u'$.

By a {\em lossy computation} of a channel system $\syst$
we mean a finite sequence:
\begin{equation}
\label{e_comp}
\trlossy{\pair{q_0}{\epsilon}}{\gamma_1}{
\trlossy{\pair{q_1}{w_1}}{\gamma_2}{
\trlossy{\pair{q_2}{w_2} \ \ \hdots \ \  }{\gamma_n}{\pair{q_n}{w_n}}}}.
\end{equation}
\begin{definition}\label{d_rea_lcs}
    {\em Lossy reachability problem} for channel systems is: given a
    channel system $S$ and a configuration $\pair{q_f}{w_f}$, with $q_f
    {\neq} q_0$, decide whether there is a lossy computation of $S$
    ending in $\pair{q_f}{w_f}$.
\end{definition}

\begin{theorem}[\cite{Sch02}]
    The lossy reachability problem for channel systems has non-primitive
    recursive complexity.
\end{theorem}

The result of~\cite{Sch02} was showed for a slightly different
model. Namely, during a single transition, a finite sequence of
messages was allowed to be read or written to the channel.
Clearly, reachability problems in both models are polynomial-time
equivalent.

In the sequel we describe a reduction from the lossy reachability for
channel systems to the emptiness problem for one-clock purely universal
alternating timed automata.  Given a channel system $\syst = (Q, q_0,
\Sigma, \Delta)$, and a configuration $\pair{q_f}{w_f}$, we
effectively construct a purely universal automaton $\autom$
with a single clock $\x$, and the input alphabet $\Sig = Q \cup \Sigma
\cup \Delta$.  The construction will assure that $\autom$ accepts
precisely correct encodings of lossy computations of $\syst$ ending in
$\pair{q_f}{w_f}$. A computation as in~(\ref{e_comp}) will be encoded
as the following word over $\Sig$:
\begin{equation}
\label{e_encoin}
q_n \gamma_n w_n \ \  q_{n{-}1} \gamma_{n{-}1} w_{n{-}1} \ \ \ldots \ \
q_1 \gamma_1 w_1 \ \  q_0,
\end{equation}
where $q_i \in Q$, $\gamma_i \in \Delta$,
$w_i \in \Sigma^*$.
Let $\syst$ be fixed in this section.

It will be convenient here to write timed words in a slightly
different way than before. From now on, whenever we write a word $w =
\pair{a_1}{t_1} \pair{a_2}{t_2} \ldots \pair{a_n}{t_n}$ we mean that
the letter $a_i$ appeared $t_i$ time units after the beginning of the
word.  In particular, $a_{i{+}1}$ appeared $t_{i{+}1} - t_i$ time
units after $a_i$.  Clearly this is correct only when $t_{i{+}1} \geq
t_i$, for $i = 1 \ldots n{-}1$.

Before the formal definition of encoding of a computation by a timed
word we outline shortly the underlying intuition. We will require that
the letter $q_n$ appears at time $0$ and then that each letter $q_i$
appears at time $n-i$. Hence, each configuration will be placed in a
unit interval. To ensure consistency of the channel contents at
consecutive configurations we require that if a message survived
during a step $i$ (it was neither read nor written nor lost) then the
distance in time between its appearances in the sequences $w_i$ and
$w_{i{-}1}$ should be precisely $1$.

We will need a new piece of notation : by $\incone{w}$ we mean the
word obtained from $w$ by increasing all $t_i$ by one time unit, i.e.,
$\incone{w} = \pair{a_1}{t_1 + 1} \pair{a_2}{t_2 + 1} \ldots
\pair{a_n}{t_n + 1}.$
\begin{definition}\label{d_enco}
    By a {\em lossy computation encoding ending in $(q_f,w_f)$} we mean
    any timed word over $\Sig$ of the form:
\begin{equation*}
    \pair{q_n}{t_n} \pair{\gamma_n}{t'_n}  v_n \
    \pair{q_{n{-}1}}{t_{n{-}1}} \pair{\gamma_{n{-}1}}{t'_{n{-}1}}
    v_{n{-}1} \ \ \ldots \ \
    \pair{q_1}{t_1} \pair{\gamma_1}{t'_1}  v_1 \
    \pair{q_0}{t_0},
\end{equation*}
where each $v_i=\pair{a^1_i}{\ti^1_i} \ \ldots \
\pair{a^{l_i}_i}{\ti^{l_i}_i}$ is a timed word over $\Sigma$.
Additionally we require that for each $i\leq n$ and $j = 1, \ldots,
l_i$,
the following conditions hold:
\begin{description}
\item[(P1)] Structure:
    \begin{equation*}
q_i \in Q, \gamma_i \in \Delta, a^j_i \in \Sigma,
\gamma_i = \langle q_{i{-}1}, x, q_i \rangle,
q_n = q_f \text{ and } a^1_n \ldots a^{l_n}_n = w_f.
    \end{equation*}

\item[(P2)] Distribution in time:
    \begin{equation*}
      n{-}i = t_i < t'_i < u^1_i < u^2_i < \ldots < u^{l_i}_i <
      t_{i{+}1} = n{-}i{+}1.
    \end{equation*}
\item[(P3a)] Epsilon move:
if $\gamma_i = \langle q_{i{-}1}, \epsilon, q_i \rangle$ then
$\incone{v_i} \subse v_{i{-}1}$.
\item[(P3b)] Write move:
if $\gamma_i = \langle q_{i{-}1}, \wri{a}, q_i \rangle$
then either $v_i=(a,u^1_i)v'$ and $v'+1\subse
v_{i{-}1}$, or $\incone{v_i} \subse v_{i{-}1}$.
\item[(P3c)] Read move:
if $\gamma_i = \langle q_{i{-}1}, \rea{a}, q_i \rangle$
then $v_{i{-}1} = v' \pair{a}{t} v''$ for some timed words $v', v''$
and $t
\in \rplus$,  such that $\incone{v_i} \subse v'$.
\end{description}
\end{definition}
\begin{lemma}\label{lemma:def}
    $\syst$ has a computation of the form~(\ref{e_comp}) ending in
    $\pair{q_n}{w_n} = \pair{q_f}{w_f}$ if and only if there exists a
    lossy computation encoding ending in $\pair{q_f}{w_f}$ as in
    Definition~\ref{d_enco}.
\end{lemma}
Our aim is:
\begin{lemma}\label{lemma:auto}
    A purely universal automaton $\Aa$ can be effectively constructed
    such that $\lang$ contains precisely all lossy computation encodings
    ending in $(q_f,w_f)$.
\end{lemma}

The proof of this lemma will occupy the rest of this section.
Automaton $\autom$ will be defined as a conjunction of four automata,
each responsible for some of the conditions from
Definition~\ref{d_enco}:
\[
\autom := \automstruct \ \land \ \automunit \ \land \ \automdelta
\ \land \ \automcheck.
\]
All four automata will be purely universal and will use at most one
clock.  Automaton $\automstruct$ verifies condition {\em (P1)},
automata $\automunit$ and $\automdelta$  jointly check condition {\em
    (P2)}, and $\automcheck$ enforces the most involved
conditions {\em (P3a) -- (P3c)}.

We omit an obvious definition of $\automstruct$. We also omit the
construction of the automaton $\automunit$ checking that letters from
$Q$ appear precisely at times $0, 1, \ldots, n$. Automaton
$\automdelta$ will accept a timed word iff the first letter is at time
0 and no two consecutive letters appear at the same time.  This can be
easily achieved by the following rules:
\begin{equation*}
{s_0},{\Sig},{\x = 0}\mapsto{\pair{s}{\emptyset}} \qquad
{s},{\Sig},{\x > 0}\mapsto{\pair{s}{\{\x\}}}.
\end{equation*}
with $s_0$ an initial state and both $s_0, s$ as accepting ones.
For readability of notation, when no clock is reset, as in the
first rule above, we will omit writing it explicitly.
Moreover, for conciseness, we implicitly assume that the automaton
fails to accept from a state,
if no rule is applicable in that state.

\notinconf{
This can be easily achieved by the following rules:
%
%
%

\begin{minipage}{0.5\linewidth}
\begin{eqnarray*}
\treq{s_0}{\Sig}{\x = 0}{\pair{s}{\emptyset}}
\end{eqnarray*}
\end{minipage}
\begin{minipage}{0.5\linewidth}
\begin{eqnarray*}
\treq{s}{\Sig}{\x > 0}{\pair{s}{\{\x\}}},
\end{eqnarray*}
\end{minipage}\\

\noindent
with $s_0$ an initial state and both $s_0, s$ as accepting ones.
}


The above mentioned automata are not only purely universal but also
purely existential, i.e., deterministic.  The power of universal
choice will be only used in the last automaton $\automcheck$, that
checks for correctness of each transition step of $\syst$.  While
analysing definition of $\automcheck$ we will comfortably assume that
an input word meets all conditions verified by the other automata,
otherwise the word is anyway not accepted.

The transition rules of $\automcheck$ from the initial state $s_0$
are as follows:
\begin{eqnarray*}
\treq{s_0}{q}{\ttt}{s_0 \land \pair{\qste}{\{\x\}}},
\quad\text{\ for } q \in Q \setminus \{q_0\} \\
\treq{s_0}{q_0}{\ttt}{\top}\\
\treq{s_0}{\Sigma \cup \Delta}{\ttt}{s_0}.
\end{eqnarray*}
Intuitively, at each $q \in Q$, except at $q_0$, an extra automaton is
run from the state $\qste$, in order to check correctness of a single
step.  Symbol $\top$ on the right-hand side stands for
a distinguished state that accepts unconditionally.

Now the rules
$\tr{\qste}{\gamma}{\ldots}{\ldots}$
depend on $\gamma = \langle q, x, q'\rangle$. There are three cases,
corresponding to conditions {\em (P3a), (P3b)} and {\em (P3c)},
respectively.
\notinfull{
Case {\em (P3b)}, not much different from {\em (P3a)}, is omitted
here.
}

\paragraph{I. Case $\g=\struct{q,\e,q'}$:}\qquad
$\tr{\qste}{\struct{q,\e,q'}}{\ttt}{\qchannel}$.\vspace{2ex}

In state $\qchannel$, the automaton checks the condition (P3a), i.e.,
whether all consecutive letters from $\Sigma$ are copied one time unit
later.  This is done by:
\begin{eqnarray*}
\treq{\qchannel}{a}{\ttt}{\qchannel \land \pair{\qpone{a}}{\{\x\}}},
\text{\ \ for } a \in \Sigma \\
\treq{\qchannel}{q}{\ttt}{\top}, \qquad \text{for $q\in Q$}.
\end{eqnarray*}
Hence, the automaton starts a check from $s^{+1}_a$ at every letter
read. Note that this is precisely here where the universal branching
is essential. The task of $\qpone{a}$ is to check that there is letter
$a$ one time unit later:
\begin{eqnarray*}
\treq{\qpone{a}}{a}{x = 1}{\top} \\
\treq{\qpone{a}}{\Sig}{x < 1}{\qpone{a}}.
\end{eqnarray*}

\paragraph{II. Case $\g = \struct{q,\wri{a},q'}$:}\qquad
$\tr{\qste}{\struct{q,\wri{a},q'}}{\ttt}{\qwri{a}}$.\vspace{2ex}

 From state $\qwri{a}$ the automaton is responsible for checking the
correctness of the operation $\wri{a}$, i.e., condition (P3b):
\begin{eqnarray*}
\treq{\qwri{a}}{a}{\ttt}{\qchannel} \\
\treq{\qwri{a}}{b}{\ttt}{\pair{\qpone{b}}{\{\x\}} \land \qchannel},
\text{\ \ for } b \in \Sigma \setminus \{a\} \\
\treq{\qwri{a}}{q}{\ttt}{\top}, \text{\ \ for } q \in Q.
\end{eqnarray*}
First rule reads simply the letter $a$ and then starts the check from
$\qchannel$. This is the correct behaviour both when the written
message was
not forgotten, and  when after forgetting it the first message is
still $a$. The second rule deals with the case when
the $a$ written to the channel has been lost immediately.  The last
rule deals with the case when not only the $a$ has been lost, but
moreover the channel is empty.

\paragraph{III. Case $\g=\struct{q,\rea{a},q'}$:}\qquad
$\tr{\qste}{\struct{q,\rea{a},q'}}{\ttt}{\qrea{a} \land
    \pair{\qtryrea{a}}{\{\x\}}}$.\vspace{2ex}

The behaviour of $\qrea{a}$ is very similar to $\qchannel$ but
additionally it will start a new copy of the automaton in the state
$\qtryrea{a}$. The goal of $\qtryrea{a}$ is to check for the letter
$a$ at the end of the present configuration.
\begin{eqnarray*}
\treq{\qrea{a}}{b}{\ttt}
{\qrea{a} \land \pair{\qpone{b}}{\{\x\}} \land \pair{\qtryrea{a}}{\{\x
\}}},
\text{\ \ for } b \in \Sigma \\
\treq{\qrea{a}}{Q}{\ttt}{\top}.
\end{eqnarray*}
Note the clock reset when entering to $\qtryrea{a}$. As we cannot know
when the configuration ends we start $\qtryrea{a}$ at each letter read.
If we realize that this was not the end (we see another channel
letter) then the check just succeeds.  If this was the end (we see a
state) then  the true check starts from the state
$\qchrea{a}$:
\begin{eqnarray*}
\treq{\qtryrea{a}}{\Sigma}{\ttt}{\top} \\
\treq{\qtryrea{a}}{Q}{\ttt}{\qchrea{a}}.
\end{eqnarray*}
 From $\qchrea{a}$ we look for some $a$
that appears more than one time unit later:
\begin{eqnarray*}
\treq{\qchrea{a}}{\Sig}{x \leq 1}{\qchrea{a}} \\
\treq{\qchrea{a}}{a}{x > 1}{\top} \\
\treq{\qchrea{a}}{b}{x > 1}{\qchrea{a}}, \text{\ \ for } b \in \Sigma
{\setminus} \{a\}.
\end{eqnarray*}
\notinconf{
The first rule skips a prefix of the following configuration. The
second one
checks successfully and accepts. The last one does not succeed but
continues to search for $a$ later.
}
Automaton $\automcheck$ has no other accepting states but $\top$.

By the very construction, $\autom$ satisfies Lemma~\ref{lemma:auto}.
By Lemma~\ref{lemma:def}, $\syst$ has a computation~(\ref{e_comp})
ending in $\pair{q_f}{w_f}$ if and only if $\lang$ is nonempty.
This completes the proof of Theorem~\ref{t_lb}.

\section{Silent transitions}\label{sec:epsilon}

In this section we point out that by extending the alternating timed  
automata model
with $\epsilon$-transitions we lose decidability. It is known that
$\epsilon$-transitions extend the power of nondeterministic timed
automata~\cite{AluDil94,BDGP98}. Here we show some evidence that every
extension of alternating timed automata with $\epsilon$-transitions
will have undecidable emptiness problem.

It turns out that there are many possible ways of introducing
$\epsilon$-transitions to alternating timed automata. To see the
issues involved consider the question of whether such an automaton
should be allowed to start uncountably many copies of itself or not.
Facing these problems we have decided not discuss virtues of different
possible definitions but rather to show where the
problem is. We will show that the universality problem for purely
existential automata with a very simple notion of
$\epsilon$-transitions is undecidable.

Timed words are written here in the same convention as in
previous section: $w = \pair{a_1}{t_1} \pair{a_2}{t_2} \ldots
\pair{a_n}{t_n}$ means that the letter $a_i$ appeared at time $t_i$
since the beginning of the computation.

We consider purely existential (i.e.\ nondeterministic) automata with
one clock.  We equip them now with additional $\epsilon$-transitions
of the form $\tr{q}{\epsilon}{\sigma}{b}$. The following trick
is used to shorten formal definitions.
\begin{definition}
    A {\em nondeterministic timed automaton with $\epsilon$-transitions}
    over
    $\Sigma$ is a nondeterministic timed automaton over
    the alphabet $\Sigeps = \Sigma \cup \{ \epsilon \}$.
\end{definition}
For convenience, we want to distinguish an automaton $\autom$ with
$\epsilon$-transitions over $\Sigma$ from the corresponding automaton
over $\Sigeps$; the latter will be denoted $\automeps$.  Given a timed
word $v$ over $\Sigeps$, by $\remeps{v}$ we mean the timed word over
$\Sigma$ obtained from $w$ by erasing all (timed) occurrences of
$\epsilon$.
\begin{definition}
    A timed word over $\Sigma$ is accepted by a timed automaton $\autom$
    with $\epsilon$-transitions if there is a timed word $v$ over
    $\Sigeps$ accepted by $\automeps$ such that $w = \remeps{v}$.
\end{definition}
Note that according to the definition, an accepting run
is always finite. The main result of this section is:
\begin{theorem}\label{thm:undec}
      The universality problem for one-clock nondeterministic timed
      automata with $\epsilon$-transitions is undecidable.
\end{theorem}
The proof is by reduction of the reachability problem for perfect
channel systems, defined similarly as lossy reachability in
Definition~\ref{d_rea_lcs}, but w.r.t.\ {\em perfect computation} of
channel systems.  Not surprisingly, a perfect computation is any
finite sequence of (perfect) transitions:
\[
\trperf{\pair{q_0}{\epsilon}}{\gamma_1}{
\trperf{\pair{q_1}{w_1}}{\gamma_2}{
\trperf{\pair{q_2}{w_2} \ \ \hdots \ \  }{\gamma_n}{\pair{q_n}{w_n}}}},
\]
%
%
\begin{theorem}[\cite{BZ83}]
    The perfect reachability problem for channel systems is undecidable,
    assumed $|\Sigma| \geq 2$.
\end{theorem}
%
%
Given a channel system $\syst
= (Q, q_0, \Sigma, \Delta)$ and a configuration $\pair{q_f}{w_f}$, we
effectively construct a one-clock nondeterministic timed automaton
with $\epsilon$-transitions $\automdru$ over $\Sig$.  Automaton
$\automdru$ will accept precisely the complement of the set of all {\em
    perfect computation encodings ending in $(q_f,w_f)$}, defined by:
\begin{definition}
\label{d_perfenco}
A {\em perfect computation encoding ending in $(q_f,w_f)$} is defined
as in Definition~\ref{d_enco}, but with the conditions {\em (P3a)} --
{\em (P3c)} replaced by:
\begin{description}
\item[(P3a)]
if $\gamma_i = \langle q_{i{-}1}, \epsilon, q_i \rangle$ then
$\incone{v_i} = v_{i{-}1}$,
\item[(P3b)]
if $\gamma_i = \langle q_{i{-}1}, \wri{a}, q_i \rangle$ then
$\incone{v_i} = \pair{a}{t} v_{i{-}1}$,
for some $t \in \rplus$.
\item[(P3c)]
if $\gamma_i = \langle q_{i{-}1}, \rea{a}, q_i \rangle$ then
$\incone{v_i \pair{a}{t}} = v_{i{-}1}$,
for some $t \in \rplus$.
\end{description}
\end{definition}
Since each perfect computation encoding is a lossy one,
$\automdru$ will be defined as a disjunction,
$ 
\automdru \ := \ \compl{\autom} \ \lor \ \automextra,
$ 
of the complement of the automaton $\autom$ from the previous section
and another automaton $\automextra$.  As automaton $\lnot \Aa$ takes
care of all timed words that are not lossy computation encodings, it
is enough to have:
\begin{lemma}\label{lemma:extra}
    Automaton $\automextra$ accepts precisely these lossy
    computation encodings ending in $(q_f,w_f)$
    that are not perfect computation encodings.
\end{lemma}
This will be enough for correctness of our reduction: $\automdru$ will
accept precisely the complement of the set of all perfect computation
encodings.

In the rest of this section we sketch the
construction of the automaton required by
Lemma~\ref{lemma:extra}.

When defining the behaviour of $\automextra$ we can conveniently
assume that the input word is already a lossy computation encoding.
The aim of $\automextra$ is to find a loss of a message in the
channel.  This will be achieved, roughly, via an $\epsilon$-rule
trying to guess a moment $t$ in time such that there is no message
occurrence at time $t$ but there is one at time $t{+}1$.
Of course, $\automextra$ (and hence $\automdru$ as well)
will have a single clock $\x$ and
the input alphabet is $\Sig = Q \cup \Sigma \cup \Delta$.

The transition rules of $\automextra$
from the initial state $s_0$ are:
\begin{eqnarray*}
\treq{s_0}{q}{\ttt}{s_0 \lor \qste}
\text{\ \ for } q \in Q \setminus \{q_0\} \\
\treq{s_0}{\Sigma \cup \Delta}{\ttt}{s_0}.
\end{eqnarray*}
Intuitively, at each $q \in Q$, except at $q_0$,
$\automextra$ chooses either to
check correctness of this single step or to skip it.
$\automextra$ will have no accepting states but $\top$ that we will
use later.

Now the rules $\tr{\qste}{\gamma}{\ldots}{\ldots}$ for state $\qste$
depend on $\gamma = \langle q, x, q'\rangle$. There are three cases,
corresponding to conditions {\em (P3a), (P3b)} and {\em (P3c)},
respectively.
As the rules follow a similar pattern to that in
Section~\ref{sec:lower}, we present
only the simplest case when
$\g=\struct{q,\epsilon,q'}$.

\begin{eqnarray*}
\treq{\qste}{\struct{q,\epsilon,q'}}{\ttt}{\pair{\qchannel}{\{\x\}}}.
\end{eqnarray*}
In state $\qchannel$, the automaton searches for a message loss.
Here we need $\epsilon$-transitions to choose the right moment
to move to state $\qone$:
\begin{eqnarray*}
\treq{\qchannel}{\epsilon}{\x > 0}{\pair{\qone}{\{\x\}}} \\
\treq{\qchannel}{\Sigma}{\ttt}{\pair{\qchannel}{\{\x\}}}
\end{eqnarray*}
The task in state $\qone$ is to wait precisely one time
unit and then check for a letter, similarly as state $\qpone{a}$ in
Section~\ref{sec:lower}.
Transition from $\qchannel$ to $\qone$ is only possible when
$\x > 0$.
As $\x$ is reset at each letter read, this ensures a positive
delay between any letter and an $\epsilon$-move.
\begin{eqnarray*}
\treq{\qone}{\Sig}{0 < \x < 1}{\qone} \\
\treq{\qone}{\Sig}{\x = 1}{\top}
\end{eqnarray*}
The only way of accepting from $\qone$ is to consume a number of letters
while $0 < \x < 1$ and finally find a letter at $\x = 1$.
Note strictness of the left-hand side inequality in $0 < \x < 1$.
It is crucial here and excludes $x = 0$, that would mean that some
letter occurred in the input word
at the moment of the $\epsilon$-move that entered into $\qone$.

This completes our description of the construction of the automaton
$\automextra$ as required by Lemma~\ref{lemma:extra}. Having it
we have the automaton $\Aa'$ which shows Theorem~\ref{thm:undec}.


\section{Infinite words}
\label{s:infin}

In this section we consider one-clock alternating timed automata over
infinite
words with B\"uchi acceptance condition.
The acceptance game is defined similarly as in Section~\ref{s:ata},
but it is played over an $\omega$-word
\[
(a_1, t_1) (a_2, t_2) \ldots,
\]
where $t_1 < t_2 < \ldots$.
Hence each play $(q_0, \val_0), (q_1, \val_1), \ldots$ is infinite.
The winner is Eve iff an accepting state appears infinitely often,
i.e., $q_i \in F$ for infinitely many indices $i$.
We do not explain the details since we will only consider
nondeterministic automata in this section (i.e., only Eva plays).
We prove the following result.

%
\begin{theorem}\label{thm:undec_Buchi}
The universality problem for one-clock nondeterministic B\"uchi
timed automata is undecidable.
\end{theorem}
As a direct corollary, emptiness problem of one-clock alternating 
B\"uchi automata is undecidable as well.

To prove Theorem~\ref{thm:undec_Buchi} we code the halting
problem of a Turing machine. We can assume that the Turing machine  
starts the empty
tape and accepts by reaching a unique accepting state $\qacc$.
Furthermore, we assume that the machine is
deterministic, i.e., we have a transition function $\delta$
specifying for each control state $q$ and tape symbol $a$
a triple $\delta(q, a) = (d, q', b)$ consisting of a head
direction $d \in \{ \leftarrow, \cdot, \rightarrow \}$, new state
$q'$ and letter $b$ to be written onto the tape in place of $a$.

The idea of the reduction is based on the fact that instead
of considering a computation that just stops in an accepting state we
will encode existence of a computation that after reaching an accepting
state clears the tape with blanks and restarts. Thus the accepting
computation is rather a repetitive accepting computation.
As the machine is deterministic, the same execution will be
essentially replayed infinitely often.

We code a sequence of configurations as before, each
configuration should fit in a unit interval. We make our simulation in
such a way that the first configuration is already of length sufficient
for the whole computation, hence in the simulation of machine steps we
will never have to add or remove tape positions.

The nondeterministic automaton we are going to construct will accept the
sequences that are not encodings of the repetitive accepting
computation of the machine. With one clock we can check that there is
a cheating, i.e., letter $a$ in one configuration is changed to $b$ in
the next although it should have not. We can also check that a letter
disappeared (it was in one configuration and not in the next).  What
we cannot check directly is that there are new letters in the next
configuration, i.e., there can appear new tape positions that were not
there before. But if this kind of inserts happen infinitely often then
we can find a sequence of tape symbols appearing at times $t_1 < t_2 <
\ldots$ such that the sequence $\fract{t_1}$, $\fract{t_2}$, $\ldots$
is either strictly increasing or strictly decreasing.  This can be
checked by a nondeterministic B\"uchi automaton with one clock.
Hence, we can construct an automaton that does not accept the  
sequences where there
are no cheatings, no disappearances and only finitely many inserts. In
such a sequence we have, from some position on, a correct and accepting
computation of the Turing machine. Thus, the nondeterministic
automaton will not accept some word iff the machine halts, i.e.,
accepts from the empty tape.

Now we will make all these intuitions more formal.
Let $\mach$ be a fixed Turing machine in the rest of this section;
by $Q$ and $\Sigma$ let us denote the set of control states and tape
alphabet of $\mach$, respectively.
Assume that a blank symbol $\bla$ is in $\Sigma$.
Given $\mach$, we will effectively construct a nondeterministic B\"uchi
automaton $\autom$ with a single clock $x$ over the input alphabet
$\Sig = Q \cup \Sigma \cup \Sigma {\times} \{\hea\}$.
A letter $\langle a, \hea \rangle$, for $a \in \Sigma$, represents a
tape symbol $a$
with the head over it.
We put $\SigH = \Sigma \cup \Sigma {\times} \{\hea\}$.

The configuration of $\mach$ is a pair $(q, w)$ consisting of a
control state
$q \in Q$ and a word $w \in \SigH^*$
representing the tape content.
The transition function $\delta$ of $\mach$ gives rise to a relation
between configurations, describing the single step of $\mach$.
We will denote this by $\move{q}{w}{q'}{w'}$, to say that a
single step from configuration $(q, w)$ yields a new configuration
$(q', w')$ and
that $w$ and $w'$ are of the same length.
So we will model computation that does not go outside $w$ with the
idea that enough space was allocated in the initial configuration.

This notation assumes a fixed size of tape available,
i.e., $w$ and $w'$ are of the same length and the head may not move
outside $w$.
For convenience, we will also write $\timedmove{q}{v}{q'}{v'}$
for \emph{timed} words $v$ and $v'$ if
$\move{q}{\ \untime{v}}{q'}{\untime{v'}}$ holds and time-stamps are
identical in $v$ and $v'$ (note that $v$ and $v'$ are of the same
length in
particular);
$\untime{v}$ stands for the word $v$ after removing time-stamps.

\begin{definition}\label{d_enco_TM}
    By a {\em recurrent accepting computation encoding} we mean
    any timed word $w$ over $\Sig$ of the form:
\begin{equation*}
    \pair{q_0}{t_0}  \     v_0 \
    \pair{q_{1}}{t_{1}} \   v_{1} \ \ \ldots,
\end{equation*}
such that the following conditions hold:
\begin{description}

\item[(P1)] Structure:
    each $q_i \in Q$ and each $v_i=\pair{a^1_i}{\ti^1_i} \ \ldots \
    \pair{a^{l_i}_i}{\ti^{l_i}_i}$ is a nonempty finite timed word
over $\SigH$
    such that precisely one of $a^1_i \ldots a^{l_i}_i$ is in $\Sigma
{\times} \{\hea\}$.

\item[(P2)] Distribution in time:
$
      i = t_i < u^1_i < u^2_i < \ldots < u^{l_i}_i <
      t_{i{+}1} = i{+}1.
$
\item[(P3)] Acceptance:
    $q_0$ is the initial state of $\mach$, each of $a^1_0 \ldots a^
{l_0}_0$ is in
    $\{ \bla, \langle \bla, \hea \rangle \}$, and $q_i = \qacc$ for
infinitely many $i$.

\item[(P4)] Recurrence: whenever $q_{i-1} = \qacc$, then $q_{i} = q_0$
    and $a^1_{i}, \ldots, a^{l_{i}}_{i} \in \{ \bla, \langle \bla,
\hea \rangle \}$.

\item[(P5)] Steps: whenever $q_{i-1} \neq \qacc$, $\timedmove{q_{i-1}}
{\incone{v_{i-1}}}{q_i}{v}$,
    for some $v \subse v_{i}$.

\item[(P6)] Insertions bound: $w$ contains no infinite subsequence $
(a_0, u_0) (a_1, u_1)
    \ldots$ such that $u_0 < u_1 < \ldots$, $a_i \in \SigH$ for all $i
\geq 0$, and the sequence
\[
\fract{u_0}, \fract{u_1}, \ldots
\]
is
    either strictly increasing or strictly decreasing.
\end{description}
\end{definition}
\begin{lemma}\label{lemma:def_MT}
    Started with the empty tape, the machine $\mach$ accepts if and only
    if there exists a recurrent accepting computation encoding as in
    Definition~\ref{d_enco_TM}.
\end{lemma}
\begin{proof}
    Assume $\mach$ accepts. There is a sequence
\[
\move{q_0}{w_0}{q_1}{w_1} \ldots
\tomove{q_n}{w_n}
\]
where $q_n = \qacc$ and $w_0$ is a finite word over
$\SigH$ representing a sufficiently big portion of initially empty
tape to store the computation.
Hence,
there is a recurrent accepting computation encoding obtained by
repeating infinitely the word $q_0 w_0 q_1 w_1 \ldots q_n w_n$;
time-stamps for tape symbols in $w_0$, $w_1$, $\ldots$ can be chosen
arbitrarily to satisfy \emph{(P2)} and \emph{(P5)}.

For the opposite direction, assume that some recurrent accepting
computation encoding $w$ exists.

By \emph{(P6)}, it contains only finitely many \emph{insertions},
where by an insertion we mean a
pair $(a, t)$, $a \in \SigH$, appearing in $w$ such that no letter
appears at time $t-1$ in $w$.
Indeed, assume otherwise, i.e., assume that the number of insertions
in $w$ is infinite.
Build the infinite sequence of all the insertions, in the
order they appear in $w$.
The fractional parts $\fract{t}$ of all the time-stamps form
an infinite sequence of reals in $(0..1)$, with no number appearing
twice.
Such a sequence has necessarily a subsequence that is either
strictly increasing or strictly decreasing -- contradiction with
\emph{(P6)}.

By \emph{(P3)} and \emph{(P4)}, $w$ contains infinitely many restarts
of the machine.
This implies that there is a restart followed by no insertion any more.
Hence, from this position on, the encoding simulates
the machine faithfully and provides the halting run of the machine.
\qed
\end{proof}
%
%

The undecidability result will follow from the next lemma.

\begin{lemma}\label{lemma:auto_MT}
    A nondeterministic
    automaton $\Aa$ can be effectively constructed
    such that $\lang$ contains precisely all timed words that are not
    recurrent accepting computation encodings.
\end{lemma}

The automaton $\Aa$ is a disjunction of six automata, each
of them accepting timed words that do not satisfy one of conditions
\emph{(P1)}--\emph{(P6)}, respectively. We omit the automata for
(negation of)
\emph{(P1)}--\emph{(P4)} and focus on the other two conditions only.
While analysing the definitions we may assume conveniently that the
input word satisfies conditions
\emph{(P1)}--\emph{(P4)}.

Automaton for negation of \emph{(P5)}, in its initial state $s_0$,
at each letter $q \in Q$ read,
decides nondeterministically either to check this step, or
to keep searching for another step to check;
in the former case, it guesses a move of the head
in this step:
\begin{eqnarray*}
\treq{s_0}{q}{\ttt}{\qleft{q} \lor \qright{q} \lor \qstay{q} \lor
s_0}, \ \text{for} \  q \in Q\\
\treq{s_0}{\SigH}{\ttt}{s_0}.
\end{eqnarray*}
To show the idea, we present in detail the transition rules from state
$\qstay{q}$ only; but we omit transitions from $\qleft{q}$ and $
\qright{q}$,
as they are conceptually similar.
In state $\qstay{q}$, the automaton needs to check that the next
configuration differs from the
configuration determined by a single machine step from the current
configuration.
The automaton can check tape symbols appearing precisely one unit
later that some symbol in the current configuration; hence insertions
are pretty allowed.
\begin{eqnarray*}
\treq{\qstay{q}}{a}{\ttt}{\pair{\qpone{a}}{\{x\}} \lor \qstay{q}}, \
\text{for} \ a \in \Sigma\\
\treq{\qstay{q}}{\langle a, \hea \rangle}{\ttt}
{\pair{\qpone{\langle b, \hea \rangle}}{\{x\}} \lor \qstaycont{q'}},
\ \text{if} \ \delta(q, a) = (\cdot, q', b)\\
\treq{\qstaycont{q}}{a}{\ttt}{\pair{\qpone{a}}{\{x\}} \lor \qstaycont
{q}}\\
\treq{\qstaycont{q}}{q'}{\ttt}{\top}, \ \text{if} \ q' \neq q.
\end{eqnarray*}
Observe that the automaton fails to accept from $\qstay{q}$ if the
head move in current configuration is not '$\cdot$', i.e,
the automaton's guess has been incorrect.
The task from state $\qpone{a}$, for $a \in \SigH$, is merely to
check that the letter appearing one unit
later is \emph{not} equal to $a$, or that there is no such letter at
all:
\begin{eqnarray*}
\treq{\qpone{a}}{\Sig}{x < 1}{\qpone{a}}\\
\treq{\qpone{a}}{b}{x = 1}{\top}, \ \text{if} \ a \neq b\\
\treq{\qpone{a}}{\Sig}{x > 1}{\top}.
\end{eqnarray*}
The only accepting state is $\top$.

Now we switch to condition \emph{(P6)}.
The task is to
recognize a strictly increasing or strictly decreasing subsequence as
defined
in \emph{(P6)}, hence the automaton is a disjunction
$\Aa_\text{inc} \lor \Aa_\text{dec}$.
For simplicity of analysis, assume that the input word satisfies all
previous conditions \emph{(P1)}--\emph{(P5)}.
In particular, for each letter appearing at time $t$, say, there is
another letter at time $t + 1$.

As a preparation, consider the following transition rules, from
states $s$ and $\bar{s}$, respectively:\\
\begin{minipage}{0.5\linewidth}
\begin{eqnarray*}
\treq{s}{\SigH}{\ttt}{s}\\
\treq{s}{Q}{x < 1}{\bar{s}}\\
\treq{s}{Q}{x = 1}{\pair{s}{\{x\}}}\\
\end{eqnarray*}
\end{minipage}
\begin{minipage}{0.5\linewidth}
\begin{eqnarray*}
\treq{\bar{s}}{\SigH}{x < 1}{\bar{s}}\\
\treq{\bar{s}}{\SigH}{x = 1}{\pair{s}{\{x\}}}\\
\treq{\bar{s}}{Q}{\ttt}{\pair{\bar{s}}{\{x\}}}\\
\end{eqnarray*}
\end{minipage}
Imagine that the clock $x$ has been reset at some letter $a \in \SigH$
of the input word.  Now, starting from state $s$, the above rules
describe scanning of the word in the following cycle: scan all letters
in $\SigH$ staying in state $s$, then on $q \in Q$ change the state to
$\bar{s}$; then scan the following letters in $\SigH$ until $x = 1$,
i.e., until precisely one time unit elapses since the last clock
reset; then reset the clock again and change to state $s$; and so on.
Hence, the whole word is conceptually split into segments determined
by the clock resets, and each segment is typically scanned in two
``phases'': first the $s$-phase and then the $\bar{s}$-phase.
The transition from $s$ to $\bar{s}$ can happen when we see a
state from $Q$; thus only at integer times by property (P2).  The only
small difference appears when one of the phases starts by a clock  
reset at
some letter $q \in Q$; in this case the other phase is degenerate and
the bottom-most transition rules for $s$ and $\bar{s}$ apply.  In fact
this is the case initially, since for the initial state of
$\Aa_\text{inc}$ and $\Aa_\text{dec}$ we choose $s$ and $\bar{s}$,
respectively.

Having these rules, definition of $\Aa_\text{inc}$ and
$\Aa_\text{dec}$ requires only appropriate handling of moments
where additional clock resets may be done.
In $\Aa_\text{inc}$ the additional clock resets will be enabled only
during $s$-phase, while in $\Aa_\text{dec}$ only in $\bar{s}$-phase.

We will need a third state $s'$ with the following rules:
\begin{eqnarray*}
\treq{s'}{\SigH}{\ttt}{s'}\\
\treq{s'}{Q}{\ttt}{\bar{s}},
\end{eqnarray*}
enabling to mimic the $s$-phase, but not enabling for any additional
clock reset until some $q \in Q$ is observed.
State $s'$ will be  the only accepting state in both $\Aa_\text{inc}$
and $\Aa_\text{dec}$ and will be intentionally visited at each
consecutive letter belonging to a strictly increasing (or
decreasing) subsequence.
Now, to complete the definition of $\Aa_\text{inc}$, we allow the
transition from $s$ to $s'$ by replacing the first rule for $s$ by
the following rule:
\begin{eqnarray*}
\treq{s}{\SigH}{\ttt}{s \lor \pair{s'}{\{x\}}}.
\end{eqnarray*}
Notice that we do not allow to reset clock more than once in one
$s$-phase (by the first rule for $s'$).
But as we have assumed \emph{(P1)}--\emph{(P5)}, we know that
each letter reappears, perhaps not identically, one unit later.
Hence we will not miss a strictly increasing subsequence,
but only ``postpone'' capturing its next element to the next
$s$-phase.

Similarly, to complete the definition of $\Aa_\text{dec}$,
we allow the transition from $\bar{s}$ to $s'$ by replacing the
first rule for $\bar{s}$ by the following one:
\begin{eqnarray*}
\treq{\bar{s}}{\SigH}{x < 1}{\bar{s} \lor \pair{s'}{\{x\}}}.
\end{eqnarray*}

This completes description of automaton $\Aa$ needed for
the proof of Lemma~\ref{lemma:auto_MT} and hence also the proof of
Theorem~\ref{thm:undec_Buchi}.



\section{Final Remarks}

In this paper we have explored the possibilities opened by the
observation that the universality problem for nondeterministic timed
automata is decidable~\cite{OW04}
We have extended this
result to obtain a class of timed automata that is closed under
boolean operations and that has decidable emptiness problem. We have
shown that despite being decidable the problem has prohibitively high
complexity. We have also considered the extension of the model with
epsilon transitions.
The undecidability result for this model
points out what makes the basic model decidable and what further
extensions are not possible. Finally, maybe somewhat surprisingly, we
prove that the universality for 1-clock nondeterministic timed
automata but over infinite words is undecidable.

We see several topics for further work:
\begin{itemize}
\item
    Adding event-clocks to the model and/or extending from timed words
to trees.
    It seems that in both cases one would still obtain a decidable  
model.
\item
    Decidability of the universality problem for one-clock co-B\"uchi
automata
    is still open.
\item
    Finding logical characterisations of the
    languages accepted by alternating timed automata with one clock.
    Since we have the closure under boolean operations, we may hope to
    find one.
\item
    Finding a different syntax that will avoid the prohibitive
    complexity of the emptiness problem. There may well be another way
    of presenting alternating timed automata that will give the same
    expressive power but for which the algorithmic complexity of the
    emptiness test will be lower.
\end{itemize}

\paragraph{Acknowledgments}
We would like to thank the referees for helpful remarks.




\end{document}